\documentclass[11pt,fleqn]{iopart}
\usepackage[utf8]{inputenc}
\usepackage[T1]{fontenc}
\usepackage{fouriernc}
\expandafter\let\csname equation*\endcsname=\relax
\expandafter\let\csname endequation*\endcsname=\relax
\usepackage{amsmath,amssymb,amsfonts,amsthm}
\usepackage{physics,units}
\usepackage{hyperref,xcolor}

\numberwithin{equation}{section}

\def\osp{{\mathfrak{osp}}}

\def\sl{{\mathfrak{sl}}}
\def\os{{\mathfrak{o}}}
\def\su{{\mathfrak{su}}}
\def\sp{{\mathfrak{sp}}}
\def\bo{{\mathcal{O}}}
\def\ji{{\mathcal{J}}}

\begin{document}

\title[The Racah algebra as a commutant and Howe duality]{The Racah algebra as a commutant and Howe duality}

\author{Julien Gaboriaud}
\ead{julien.gaboriaud@umontreal.ca}
\address{Centre de Recherches Math\'ematiques, Universit\'e de Montr\'eal, Montr\'eal (QC), Canada}

\author{Luc Vinet}
\ead{luc.vinet@umontreal.ca}
\address{Centre de Recherches Math\'ematiques, Universit\'e de Montr\'eal, Montr\'eal (QC), Canada}

\author{St\'ephane Vinet}
\ead{stephanevinet@uchicago.edu}
\address{The College, The University of Chicago, 5801 S. Ellis Ave, Chicago, IL 60637, USA}

\author{Alexei Zhedanov}
\ead{zhedanov@yahoo.com}
\address{Department of Mathematics, Renmin University of China, Beijing 100872, China}

\vspace{10pt}

\begin{abstract}   
The Racah algebra encodes the bispectrality of the eponym polynomials.  It is known to be the symmetry algebra of the generic superintegrable model on the $2$-sphere. It is further identified as the commutant of the $\os(2) \oplus \os(2) \oplus \os(2)$ subalgebra of $\os(6)$ in oscillator representations of the universal algebra of the latter. How this observation relates to the $\su(1,1)$ Racah problem and the superintegrable model on the $2$-sphere is discussed on the basis of the Howe duality associated to the pair $\big(\os(6)$, $\su(1,1)\big)$.
\end{abstract}

\section{Introduction}\label{sec_intro}
The Racah algebra $\mathcal{R}$ has three generators $K_1$, $K_2$, $K_3 $ that are subjected to the relations: 
\begin{align}\label{eq_racah}
\begin{aligned}{}
 [K_1,K_2]=K_3 \qquad\quad\  [K_2,K_3]&={K_2}^2+\{K_1,K_2\}+dK_2+e_1 \quad\ \\
 [K_3,K_1]&={K_1}^2+\{K_1,K_2\}+dK_1+e_2
\end{aligned}
\end{align}
where $[A,B] = AB-BA$, $\{A,B\}=AB+BA$ and $d$, $e_1$, $e_2$ are central. 
This algebra has appeared in many guises and we here add to its understanding with the identification of a new realization. We shall indeed show that $\mathcal{R}$ arises as the commutant of  the $\os(2) \oplus \os(2) \oplus \os(2)$ subalgebra of $\os(6)$ in the representations of $\mathcal{U}(\os(6))$ on the Hilbert space of states of six oscillators.

The Racah algebra was introduced as an encoding of the bispectrality properties of the Racah polynomials, see for example \cite{citGVZ1}. It is also intimately connected to the recouplings of $\su(2)$ and $\su(1,1)$ representations \cite{citGZ, citGVZ2} since the Racah coefficients for these Lie algebras are expressed in terms of the Racah polynomials. We shall review this last aspect in the next section because it is relevant for the results we want to present in this paper.

The Racah algebra has also been found \cite{citKMP} to be the symmetry algebra of the generic superintegrable model on the two-sphere with Hamiltonian $H$ given by 
\begin{equation}\label{eq_H}
 H=\ji_1^2+\ji_2^2+\ji_3^2+\frac{a_1}{{x_1}^2}+\frac{a_2}{{x_2}^2}+\frac{a_3}{{x_3}^2} 
\end{equation}
where 
\begin{equation}\label{eq_Jk1}
 \ji_k=\epsilon_{kij}\,\, x_i \frac{\partial}{\partial x_j}, \quad\qquad i,\,j,\,k=1,\,2,\,3,\quad\qquad {x_1}^2+{x_2}^2+{x_3}^2=1
\end{equation}
and $a_1$, $a_2$, $a_3$ are parameters. We shall also bring this fact to bear on our discussion. 

The Racah algebra has further been shown to have a natural embedding in $\su(2)$ \cite{citGWH, citGVZ3} and to be related to distance-regular graphs \cite{citLHG}. It also extends to arbitrary ranks \cite{citDGVV} with a connection to multivariate Racah polynomials of the Tratnik type \cite{citT,citGI} and higher dimensional superintegrable models \cite{citI}.  We shall here add to this the commutant realization mentioned above and explain how Howe duality \cite{citH1,citH2,citRCR} relates this observation to the fact that $\mathcal{R}$ is also the commutant in $\mathcal{U}\big(\su(1,1)^{\otimes 3}\big)$ of the addition of three $\su(1,1)$.
 
The paper will unfold as follows. As already indicated, we shall review in \Sref{sec_racah_problem} the occurrence of the Racah algebra in the recoupling of three irreducible representations of $\su(1,1)$. This is where $\mathcal{R}$ will be the commutant of $\su(1,1)$ in $\mathcal{U}\big(\su(1,1)^{\otimes 3}\big)$ with the intermediate Casimir operators as generators. Our main result will be the object of \Sref{sec_racah_so6} where the connection with the Racah algebra and the Lie algebra $\os(6)$ will be made. The link between this last incarnation of $\mathcal{R}$ as the commutant of the maximal Abelian subalgebra $\os(2) \oplus \os(2) \oplus \os(2)$ of $\os(6)$ in the universal algebra of the latter and the realization stemming from the Racah problem of $\su(1,1)$ will be discussed in \Sref{sec_racah_howe}. This will be done by considering six harmonic oscillators and the dual reductive pair $\big(\os(6), \sp(2)\big)$ in $\sp(12)$ that acts on the Hilbert space of their collective states. This Howe duality will be invoked to put in correspondance the $\os(6)$ and $\sp(2)\simeq\su(1,1)$ pictures for the Racah algebra. In the last section, we shall complete the analysis by rederiving the results pertaining to the symmetry of the generic superintegrable model on the $2$-sphere. To that end, we shall carry out the dimensional reduction of the six-dimensional oscillator under the action of $O(2) \otimes O(2) \otimes O(2)$ to obtain the Hamiltonian from the total Casimir operator for the addition of six metaplectic representations of $\sp(2)$ and the constants of motion from the proper intermediate Casimirs with the knowledge that they realize the Racah algebra.  Summary and outlook will form the Conclusion.

\section{The Racah problem for $\su(1,1)$ and the Racah algebra}\label{sec_racah_problem}
The Lie algebra $\su(1,1)$ has generators $J_0, J_\pm$ that obey the following commutation relations:
\begin{align}
 [J_0,J_\pm]=\pm J_\pm  \qquad [J_+,J_-]=-2J_0
\end{align}
and its Casimir operator is given by:
\begin{align}
C=J_0^2-J_+J_--J_0.
\end{align}
Consider now the addition of $3$ irreducible representations of $\su(1,1)$ for which the initial Casimir operators take values~  $C^{(i)}=\lambda_i$,~  $i=1,2,3$, and let us write
\begin{align}
 J_0^{(123)}=J_0^{(1)}+J_0^{(2)}+J_0^{(3)}, \qquad J_\pm^{(123)}=J_\pm^{(1)}+J_\pm^{(2)}+J_\pm^{(3)}
\end{align}
with the superindex denoting on which of the three factors in $\su(1,1)^{\otimes 3}$ the operator is acting.
In addition to the initial Casimir operator $C^{(i)}$ we also have the intermediate Casimir operators associated to the addition of two representations 
\begin{align}
 C^{(ij)}&=\big(J_0^{(i)}+J_0^{(j)}\big)^2 -\big(J_+^{(i)}+J_+^{(j)}\big)\big(J_-^{(i)}+J_-^{(j)}\big)-\big(J_0^{(i)}+J_0^{(j)}\big)
\end{align}
with $(ij)=(12),(23),(31)$ and also the total Casimir operator, given by:
\begin{align}
 C^{(123)}=\big(J_0^{(123)}\big)^2-J_+^{(123)} J_-^{(123)}-J_0^{(123)}.
\end{align}
Take $C^{(123)}=\lambda_4$.  Let $V^{(\lambda_i)}$ denote irreducible representation spaces of $\su(1,1)$, and look at the decomposition of  \mbox{$V^{(\lambda_1)}\otimes V^{(\lambda_2)}\otimes V^{(\lambda_3)}$} in irreducibles $V^{(\lambda_4)}$. The Racah problem for $\su(1,1)$  is about determining the unitary transformations between the bases corresponding to the steps ~$(1\oplus2)\oplus 3$ and~ $1\oplus(2\oplus3)$ that respectively diagonalize the intermediate Casimirs $C^{(12)}$ and $C^{(23)}$
\begin{align}
\begin{aligned}{}
 C^{(12)}&=2J_0^{(1)}J_0^{(2)}-\big(J_+^{(1)}J_-^{(2)}+J_-^{(1)}J_+^{(2)}\big)+\lambda_1+\lambda_2,\\
 C^{(23)}&=2J_0^{(2)}J_0^{(3)}-\big(J_+^{(2)}J_-^{(3)}+J_-^{(2)}J_+^{(3)}\big)+\lambda_2+\lambda_3.
\end{aligned}
\end{align}
These intermediate Casimir operators generate the Racah algebra $\mathcal{R}$, since the relations \eref{eq_racah} are satisfied by $K_1=-\frac{1}{2}C^{(12)}$ and $K_2=-\frac{1}{2}C^{(23)}$  
with~ \mbox{$d=\frac{1}{2}\big(\lambda_1+\lambda_2+\lambda_3+\lambda_4\big)$}, \mbox{$e_1=\frac{1}{4}\big(\lambda_1-\lambda_4\big)\big(\lambda_2-\lambda_3\big)$}, ~\mbox{$e_2=\frac{1}{4}\big(\lambda_1-\lambda_2\big)\big(\lambda_4-\lambda_3\big)$}.

\section{The Racah algebra and $\os(6)$}\label{sec_racah_so6}
We will now observe that $\mathcal{R}$ is the commutant in $\mathcal{U}(\os(6))$ of a subalgebra of $\os(6)$ in the oscillator representation. The algebra $\os(6)$ has $15$ generators $L_{\mu\nu}=-L_{\nu\mu}$,~ $\mu,\nu=1,...,6$ obeying 
\begin{align}\label{eq_o6_comm}
 [L_{\mu\nu},L_{\rho\sigma}]=\delta_{\nu\rho}L_{\mu\sigma}-\delta_{\nu\sigma}L_{\mu\rho}-\delta_{\mu\rho}L_{\nu\sigma}+\delta_{\mu\sigma}L_{\nu\rho}
\end{align}
and it possesses the following quadratic Casimir:
\begin{align}
 \mathcal{C}=\sum_{\mu<\nu}L_{\mu\nu}^2.
\end{align}
We will use the realization 
\begin{align}\label{eq_realization}
 L_{\mu\nu}=\xi_\mu\frac{\partial}{\partial\xi_\nu}-\xi_\nu\frac{\partial}{\partial\xi_\mu}\qquad \mu\neq \nu,\quad \mu,\nu=1,...,6.
\end{align}
Pick the \mbox{$\os(2)\oplus\os(2)\oplus\os(2)$} subalgebra of $\os(6)$ generated by the commutative set \mbox{$\{L_{12},\,L_{34},\,L_{56}\}$}. We want to focus on the commutant of this Abelian subalgebra in $\mathcal{U}(\os(6))$. It is fairly easy to see that it will be generated by the following two invariants:
\begin{align}
 K_1&=\frac{1}{8}\big(L_{12}^2+L_{34}^2+L_{13}^2+L_{23}^2+L_{14}^2+L_{24}^2\big)\\
 K_2&=\frac{1}{8}\big(L_{34}^2+L_{56}^2+L_{35}^2+L_{36}^2+L_{45}^2+L_{46}^2\big).\label{eq_K2}
\end{align}
Define $K_3$ by~ $[K_1,K_2]=K_3$. One then finds that
\begin{align}\label{eq_K3}
\begin{aligned}{}
 K_3=\frac{1}{16}\big(&L_{35}^2+L_{36}^2+L_{45}^2+L_{46}^2-L_{13}^2-L_{14}^2-L_{23}^2-L_{24}^2-L_{15}^2-L_{16}^2-L_{25}^2-L_{26}^2\\
 &+L_{13}L_{35}L_{15}+L_{13}L_{36}L_{16}+L_{23}L_{35}L_{25}+L_{23}L_{36}L_{26}+L_{14}L_{45}L_{15}\\
 &+L_{14}L_{46}L_{16}+L_{24}L_{45}L_{25}+L_{24}L_{46}L_{26}\big).
\end{aligned}
\end{align}
Working out the commutation relations of $K_3$ with $K_1$ and $K_2$, we find that they correspond to those of a central extension of the Racah algebra with $L_{12},L_{34},L_{45}$ and $\mathcal{C}$ playing role of structure constants. Indeed one obtains
\begin{align}\label{eq_result}
\hspace{-1em}\begin{aligned}{}
 [K_1,K_2]&=K_3\\
 [K_2,K_3]&=K_2^{2}+\{K_1,K_2\}-\frac{1}{8}K_2\big(\mathcal{C}+L_{12}^2+L_{34}^2+L_{56}^2\big)-\frac{1}{64}\big(\mathcal{C}-L_{12}^2-4\big)\big(L_{34}^2-L_{56}^2\big)\\
 [K_3,K_1]&=K_1^{2}+\{K_1,K_2\}-\frac{1}{8}K_2\big(\mathcal{C}+L_{12}^2+L_{34}^2+L_{56}^2\big)-\frac{1}{64}\big(\mathcal{C}-L_{56}^2-4\big)\big(L_{34}^2-L_{12}^2\big)
\end{aligned}
\end{align}
where the parameters~ \mbox{$d=-\frac{1}{8}\big(\mathcal{C}+L_{12}^2+L_{34}^2+L_{56}^2\big)$},~ \mbox{$e_1=-\frac{1}{64}\big(\mathcal{C}-L_{12}^2-4\big)\big(L_{34}^2-L_{56}^2\big)$}, \mbox{$e_2=-\frac{1}{64}\big(\mathcal{C}-L_{56}^2-4\big)\big(L_{34}^2-L_{12}^2\big)$} are obviously central.\\

Let us indicate how this result is obtained. Take for example the commutator $[K_2,K_3]$ ( $[K_3,K_1]$ is treated similarly). On the one hand, it is readily observed that the r.h.s in \eref{eq_result} only contains terms of the form $L_{\mu\nu}^{2}$. On the other hand, using the $\os(6)$ commutation relations \eref{eq_o6_comm} and the explicit expressions \eref{eq_K2}, \eref{eq_K3} for $K_2$, $K_3$, one obtains:
\begin{align}
\begin{aligned}{}\label{eq_shit}
 \hspace{-1em}128[K_2&,~K_3]=\{{L_{35}}^{2},{L_{13}}^{2}\}+\{{L_{35}}^{2},{L_{23}}^{2}\}+\{{L_{36}}^{2},{L_{13}}^{2}\}+\{{L_{36}}^{2},{L_{23}}^{2}\}\\
 &+\{{L_{45}}^{2},{L_{14}}^{2}\}+\{{L_{45}}^{2},{L_{24}}^{2}\}+\{{L_{46}}^{2},{L_{14}}^{2}\}+\{{L_{46}}^{2},{L_{24}}^{2}\}\\
 &-\{{L_{35}}^{2},{L_{15}}^{2}\}-\{{L_{35}}^{2},{L_{25}}^{2}\}-\{{L_{36}}^{2},{L_{16}}^{2}\}-\{{L_{36}}^{2},{L_{26}}^{2}\}\\
 &-\{{L_{45}}^{2},{L_{15}}^{2}\}-\{{L_{45}}^{2},{L_{25}}^{2}\}-\{{L_{46}}^{2},{L_{16}}^{2}\}-\{{L_{46}}^{2},{L_{26}}^{2}\}\\
 &+4({L_{13}}^{2}+{L_{23}}^{2}+{L_{14}}^{2}+{L_{24}}^{2}-{L_{15}}^{2}-{L_{25}}^{2}-{L_{16}}^{2}-{L_{26}}^{2})\\
 &-2L_{15}L_{35}L_{36}L_{16}-2L_{13}L_{35}L_{56}L_{16}-2L_{25}L_{35}L_{36}L_{26}-2L_{23}L_{35}L_{56}L_{26}\\
 &-2L_{16}L_{36}L_{35}L_{15}+2L_{13}L_{36}L_{56}L_{15}-2L_{26}L_{36}L_{35}L_{25}+2L_{23}L_{36}L_{56}L_{25}\\
 &-2L_{15}L_{45}L_{46}L_{16}-2L_{14}L_{45}L_{56}L_{16}-2L_{25}L_{45}L_{46}L_{26}-2L_{24}L_{45}L_{56}L_{26}\\
 &-2L_{16}L_{46}L_{45}L_{15}+2L_{14}L_{46}L_{56}L_{15}-2L_{26}L_{46}L_{45}L_{25}+2L_{24}L_{46}L_{56}L_{25}\\
 &+2L_{14}L_{45}L_{35}L_{13}-2L_{14}L_{34}L_{35}L_{15}+2L_{24}L_{45}L_{35}L_{23}-2L_{24}L_{34}L_{35}L_{25}\\
 &+2L_{14}L_{46}L_{36}L_{13}-2L_{14}L_{34}L_{36}L_{16}+2L_{24}L_{46}L_{36}L_{23}-2L_{24}L_{34}L_{36}L_{26}\\
 &+2L_{13}L_{35}L_{45}L_{14}+2L_{13}L_{34}L_{45}L_{15}+2L_{23}L_{35}L_{45}L_{24}+2L_{23}L_{34}L_{45}L_{25}\\
 &+2L_{13}L_{36}L_{46}L_{14}+2L_{13}L_{34}L_{46}L_{16}+2L_{23}L_{36}L_{46}L_{24}+2L_{23}L_{34}L_{46}L_{26}.
\end{aligned}
\end{align}
The terms of the type $L_{\mu\nu}L_{\rho\nu}L_{\rho\sigma}L_{\mu\sigma}$ thus need to be re-expressed. The key to rewriting them with factors involving only the $L_{\mu\nu}^{2}$'s is to make use of the identity 
\begin{equation}\label{eq_key}
 L_{\mu\nu}L_{\rho\sigma}+L_{\mu\rho}L_{\sigma\nu}+L_{\mu\sigma}L_{\nu\rho}=0
\end{equation}
which is directly proved in the realization \eref{eq_realization} (and which in fact remains true for Dunkl angular momenta (\cite{citFH})), and to also take its square, which yields
\begin{align}\label{eq_key2}
 \{L_{\mu\nu}L_{\rho\sigma},\ L_{\mu\rho}L_{\nu\sigma}\}=L_{\mu\nu}^{2}L_{\rho\sigma}^{2}+L_{\mu\rho}^{2}L_{\nu\sigma}^{2}-L_{\mu\sigma}^{2}L_{\nu\rho}^{2}.
\end{align}
Combining these two identities and calling upon other elementary formulas such as
\begin{align}
 [L_{\sigma\mu}^{2}+L_{\sigma\nu}^{2},\ L_{\mu\nu}]=0
\end{align}
allows one to equate \eref{eq_shit} with the r.h.s in \eref{eq_result}, which completes the proof.

\section{The Racah algebra and Howe duality}\label{sec_racah_howe}
We shall now show how the result in the previous section can be explained by identifying the Howe pair in play in the system we have considered.
In the last two sections we showed that the Racah algebra is the commutant of $\su(1,1)$ ~in~ $\mathcal{U}\big(\su(1,1)^{\otimes3}\big)$ and of {$\os(2)\oplus\os(2)\oplus\os(2)$ ~in~ $\mathcal{U}(\os(6))$. The connection between these two observations can be traced to Howe duality.

It is known (\cite{citRCR}) that $\os(n)$ and $\sp(2d)$ form a dual pair in $\sp(2dn)$, i.e. these two subgroups are mutual commutants. This implies that $\os(6)$ and $\sp(2)\simeq \su(1,1)$ have dual actions on the Hilbert space of six oscillators. That means that their irreducible representations can be paired and this can be done through the Casimirs.\\[0.5em]
Consider the following $6$ oscillator realizations of $\sp(2)$:
\begin{align}\label{eq_su11J}
 J_{+}^{(\mu)}=\frac{1}{2}\xi_\mu^2,\qquad\ J_{-}^{(\mu)}=\frac{1}{2}\frac{d^2}{d\xi_\mu^2},\qquad\ J_0^{(\mu)}=\frac{1}{2}\xi_\mu\frac{d}{d\xi_\mu}+\frac{1}{4},\qquad\ \mu=1,2,3,4,5,6.
\end{align}
We shall add these six representations by first coupling the three pairs $(\mu\nu)=(12),(34),(56)$ and shall write:
\begin{align}
 J^{(\mu\nu)}=J^{(\mu)}+J^{(\nu)}, \qquad\qquad
 J^{(123456)}=J^{(12)}+J^{(34)}+J^{(56)}.
\end{align}
Recall that the $\sp(2)$ Casimir is $C=J_0^2-J_+J_--J_0$. The connection between the Casimirs of combined metaplectic representations with elements in $\mathcal{U}(\os(6))$ is readily obtained:
\begin{align}
 C^{(\mu\nu)}&=-\frac{1}{4}\big(L_{\mu\nu}^2+1\big),\\
 C^{(123456)}&=-\frac{1}{4}\left(\sum_{\mu<\nu}L_{\mu\nu}^2-3\right)=-\frac{1}{4}\mathcal{C}+\frac{3}{4},\\
 C^{(1234)}&=-\frac{1}{4}\big(L_{12}^2+L_{34}^2+L_{13}^2+L_{23}^2+L_{14}^2+L_{24}^2\big)=-2K_1,\\
 C^{(3456)}&=-\frac{1}{4}\big(L_{34}^2+L_{56}^2+L_{35}^2+L_{36}^2+L_{45}^2+L_{46}^2\big)=-2K_2.
\end{align}
In addition to observing that $ C^{(123456)}$ and the Casimir $\mathcal{C}$ of $\os(6)$ are affinely related, we see that the intermediate $\sp(2)$ Casimirs correspond to the generators of the commutant of $\{L_{12},L_{34},L_{56}\}$ in $\mathcal{U}(\os(6))$. We know from \Sref{sec_racah_so6}, that the intermediate $\sp(2)$ Casimirs realize the commutation relations of the Racah algebra. This will hence be the case also for the commutant generators and we have here our duality connection.

\section{The Racah algebra and the generic superintegrable model on $S^{2}$}\label{sec_racah_generic_model}
We can now complete the picture by performing the dimensional reduction from $\mathbb R^6$ to $\mathbb R^+  \times S^{2}$ \cite{citGHW, citAHP, citRTW} to obtain the generic superintegrable model (introduced in \Sref{sec_intro}) with Hamiltonian $H$ and to recover as well its symmetries. Make the following change of variables: 
\begin{align}
\begin{aligned}{}
 \xi_{2i-1}&=x_i\cos{\theta_i},\\
 \xi_{2i}&=x_i\sin{\theta_i},
\end{aligned}
 \qquad L_{2i-1,2i}=\xi_{2i-1}\frac{\partial}{\partial\xi_{2i}}-\xi_{2i}\frac{\partial}{\partial\xi_{2i-1}}=\frac{\partial}{\partial\theta_i}, \qquad i=1,2,3.
\end{align}
Eliminate the ignorable $\theta_i s$ by separating these variables and setting $L_{2i-1,2i}^2\sim k_i^2$. After performing the gauge transformation $\bo\mapsto\widetilde{\bo}=x_i^{\frac{1}{2}}\,\,\bo\,\,x_i^{-\frac{1}{2}}$ one obtains
\begin{align}
 \hspace{-1em}\widetilde{J_+}^{(2i-1,2i)}=\frac{1}{2}x_i^2,\qquad \widetilde{J_-}^{(2i-1,2i)}=\frac{1}{2}\left(\frac{d^2}{dx_i^2}+\frac{a_i}{x_i^2}\right),\qquad\widetilde{J_0}^{(2i-1,2i)}=\frac{1}{2}\left(x_i\frac{d}{dx_i}+\frac{1}{2}\right),
\end{align}
with $a_i=k_i^2+\frac{1}{4}$. The reduced Casimirs
\begin{align}		
 \widetilde{C}^{(\mu,\nu, \rho, \sigma)}&=(\widetilde{J_0})^2-\widetilde{J_+}\widetilde{J_-}-\widetilde{J_0}\quad \text{with} \quad \widetilde{J}=\widetilde{J}^{(\mu, \nu)}+\widetilde{J}^{(\rho, \sigma)},
\end{align}
are easily computed and have the following expressions:
\begin{align}
\begin{aligned}{}
 \widetilde{C}^{(1234)}&=-\frac{1}{4}\left[\ji_3^2+a_1\frac{x_2^2}{x_1^2}+a_2\frac{x_1^2}{x_2^2}+a_1+a_2+1\right]\\
 \widetilde{C}^{(3456)}&=-\frac{1}{4}\left[\ji_1^2+a_2\frac{x_3^2}{x_2^2}+a_3\frac{x_2^2}{x_3^2}+a_2+a_3+1\right]\\
 \widetilde{C}^{(1256)}&=-\frac{1}{4}\left[\ji_2^2+a_3\frac{x_1^2}{x_3^2}+a_1\frac{x_3^2}{x_1^2}+a_1+a_3+1\right],\qquad\quad  \ji_k=\epsilon_{kij}\,x_i\frac{d}{dx_j}.
\end{aligned}
\end{align}
Using $\widetilde{J}^{(123456)}=\widetilde{J}^{(12)}+\widetilde{J}^{(34)}+\widetilde{J}^{(56)}$ it is seen that
\begin{align}
 \widetilde{C}^{(123456)}=\widetilde{C}^{(1234)}+\widetilde{C}^{(3456)}+\widetilde{C}^{(1256)}-\widetilde{C}^{(12)}-\widetilde{C}^{(34)}-\widetilde{C}^{(56)}.
\end{align}
Since $\widetilde{C}^{(2i-1,2i)}=-\frac{1}{4}\big(a_i+\frac{3}{4}\big)$ are constants, the invariant can be taken to be given by the sum of the first three terms in $ \widetilde{C}^{(123456)}$. Assuming \mbox{${x_1}^2+{x_2}^2+{x_3}^2=1$}, this is recognized to be the Hamiltonian \eref{eq_H} of the generic model (up to an affine transformation). The Casimirs are essentially the conserved quantities:
\begin{align}
 Q_i=\ji_i^{2}+a_j\frac{x_k^2}{x_j^2}+a_k\frac{x_j^2}{x_k^2} \qquad i,j,k \in \{1,2,3\}\quad \text{cyclic}
\end{align}
and they generate $\mathcal R$ which is hence the symmetry algebra.

\section{Conclusion}
The Racah algebra $\mathcal{R}$ embodies the theory of the Racah polynomials. The ubiquity of these orthogonal polynomials explains the diverse roles that $\mathcal{R}$ plays and motivates, with an eye to generalizations, the examination of all facets of this algebra. It is hence quite nice that we could find a new characterization of $\mathcal{R}$ . Put in simple terms, our findings can be summarized as follows. We have shown that the Racah algebra is generated by polynomials in the generalized angular momenta in six dimensions that are invariant under rotations in three non-intersecting planes.
We have further indicated that this picture is dual, in the sense of Howe, to the one where the Racah algebra is generated by the Casimir operators in the addition of three irreducible representations of $\su(1,1)$. This was done by exploiting the correspondence between the representations of $\os(6)$ and those of $\sp(2)$ acting on the state space of a six-dimensional harmonic oscillator. The analysis provided an illuminating context within which the Racah symmetry of the generic superintegrable model on the $2$-sphere is naturally  obtained by dimensional reduction. This suggests numerous potential extensions.
 
It should be possible to extend all the results of this paper to higher dimensions, namely, to Racah algebras with rank superior to one. These algebras have already been introduced using the recoupling model, that is, as the ones generated by the various Casimir operators arising the the addition of four and more $\su(1,1)$ representations \cite{citDGVV}. In view of our observations, it is natural to expect that these could be shown to be in duality with commutants of the $n$-torus in $\os(2n)$ with $n\geq{4}$. 
 
There are also two other important rank one algebras that share properties with the Racah algebra: the Askey-Wilson (AW) and the Bannai-Ito (BI) algebras. The AW algebra \cite{citZ} accounts for the bispectral properties of the Askey-Wilson polynomials. It is the object of much attention and like the Racah algebra it arises in particular in a Racah problem, this time for the quantum algebra $\mathcal{U}_q(\sl(2))$ \cite{citGZ2}. The Bannai-Ito polynomials are most simply defined as a $q\to{-1}$ limit of the Askey-Wilson polynomials \cite{citBI}. They are also bispectral and the BI algebra \cite{citTVZ} encodes these defining features. The Racah problem of relevance in this case is the one associated to the Lie superalgebra $\osp(1,2)$ leading to a realization of the BI algebra again, in terms of the intermediate Casimir operators \cite{citGVZ4}. The BI algebra has also been shown to be the symmetry algebra of a superintegrable model on the two-sphere involving reflection operators \cite{citGVZ5} as well as of a Dirac-Dunkl equation in $\mathbb R^3$ \cite{citDGV}. This is observed by realizing the three $\osp(1,2)$ that are added in terms of Dunkl operators \cite{citD} and using a Clifford algebra in the latter problem. For a review of the BI algebra and its applications see \cite{citDGTVZ}. For both the AW and BI cases, it would be quite interesting to determine if there is a Howe duality setting that would allow to develop for these algebras a commutant interpretation similar to the one that we have found for the Racah algebra. How dimensional reduction would then operate would be revealing with respect to superintegrable models. The AW and BI algebras of higher ranks could then as well lend themselves to similar descriptions. We plan to examine all these questions in the near future.

\ack
The authors wish to thank Luc Frappat, John Harnad, \'Eric Ragoucy and Paul Sorba for enlightening discussions. 
JG holds an Alexander-Graham-Bell scholarship from the Natural Science and Engineering Research Council (NSERC) of Canada. 
LV gratefully acknowledges his support from NSERC. 
SV enjoys a Neubauer No Barriers scholarship at the University of Chicago and benefitted from a Metcalf internship.
The work of AZ is supported by the National Foundation of China (Grant No. 11771015).


\vspace{0em}
\section*{References}
\begin{thebibliography}{99}
\bibitem{citGVZ1}
V. X. Genest, L. Vinet, A. Zhedanov, \emph{The Racah algebra and superintegrable models}, J. Phys. Conf. Ser. {\bf512} (2014), 012011 

\bibitem{citKLS} 
R. Koekoek, P.A. Lesky, and R.F. Swarttouw, \emph{Hypergeometric orthogonal polynomials and their q-analogues}. Springer, 1-st edition, 2010

\bibitem{citGZ}
Y. Granovski, A. Zhedanov, \emph{Nature of the symmetry group of the 6-j symbol}, Soviet Physics JETP {\bf67} (1988), 1982-1985

\bibitem{citGVZ2}
V. X. Genest, L. Vinet, A. Zhedanov, \emph{Superintegrability in two dimensions and the Racah-Wilson algebra}, Lett. Math. Phys. {\bf104} (2014), 931-952

\bibitem{citKMP}
E.G. Kalnins, W. Miller Jr., S. Post,  \emph{Wilson polynomials and the generic superintegrable system on the 2-sphere}, J. Phys. A: Math. Theor. {\bf40} (2007), 11525-11538

\bibitem{citGWH}
S. Gao, Y. Wang, B. Hou, \emph{The classification of Leonard triples of Racah type}, Lin. Alg. Appl, {\bf439} (2013), 1834

\bibitem{citGVZ3}
V. X. Genest, L. Vinet, A. Zhedanov,  \emph{The equitable Racah algebra from three $\su(1,1)$ algebras}, J. Phys. A: Math. Theor. {\bf47} (2014), 025203

\bibitem{citLHG}
H. Liu, B. Hou, S. Gao, \emph{Leonard triples, the Racah algebra and some distance-regular graphs of Racah type}, Lin. Alg. Appl. {\bf484} (2015), 435

\bibitem{citDGVV}
H. De Bie, V. X. Genest, W. van de Vijver, L. Vinet, \emph{A higher rank Racah algebra and the $\mathbb Z_2^n$ Laplace-Dunkl operator}, J.Phys. A: Math. Theor. {\bf51} (2017), 

\bibitem{citT}
M. V. Tratnik, \emph{Some multivariable orthogonal polynomials of the Askey tableau-discrete families}, J. Math. Phys. {\bf32} (1991), 2337-2342

\bibitem{citGI}
J. Geronimo, P. Iliev, \emph{Bispectrality of multivariable Racah-Wilson polynomials}, Constr. Approx. {\bf31} (2010), 417-457

\bibitem{citI}
P. Iliev, \emph{The generic quantum superintegrable system on the sphere and Racah operators}, Lett. Math. Phys. {\bf107} (2017), 2029-2045

\bibitem{citH1}
R. Howe, \emph{Remarks on classical invariant theory}, Trans. Am. Math. Soc. {\bf313} (1989), 539

\bibitem{citH2}
R. Howe, \emph{Dual pairs in physics: harmonic oscillators, photons, electrons, singletons} in: Applications of Group Theory in Physics and Mathematical Physics, M. Flato, P. Sally, G. Zuckerman (eds.), Lectures in Applied Mathematics {\bf21}, AMS, Providence, R.I. (1985), 179-206

\bibitem{citRCR}
D. J. Rowe, M. J. Carvalho, J. Repka, \emph{Dual pairing of symmetry and dynamical groups in physics}, Rev. Mod. Phys. {\bf84} (2012), 711

\bibitem{citFH} 
M. Feigin and T. Hakobyan, \emph{On Dunkl angular momenta algebra}, J. High. Energy. Phys. {\bf 11} (2015), 107.

\bibitem{citGHW}
L. Gagnon, J. Harnad, J. Hurtubise, P. Winternitz, \emph{Abelian integrals and the reduction method for an integrable Hamiltonian system}, J. Math. Phys. {\bf26} (1985), 1605.

\bibitem{citAHP}
M. R. Adams, J. Harnad, E. Previato, \emph{Isospectral Hamiltonian Flows in Finite and Infinite Dimensions}, Comm. Math. Phys. {\bf117} (1988) 451.

\bibitem{citRTW}
M. A. Rodr\'iguez, P. Tempesta, P. Winternitz, \emph{Reduction of superintegrable systems: The anisotropic harmonic oscillator}, Phys. Rev. E {\bf78} (2008), 046608.

\bibitem{citZ}
A.S. Zhedanov, \emph{Hidden symmetry of Askey-Wilson polynomials}, Teoret. Mat. Fiz. {\bf  89} (1991) 190-204.

\bibitem{citGZ2}
Y. Granovski, A. Zhedanov, \emph{Hidden symmetry of the Racah and Clebsch-Gordan Problems for the quantum algebra $\sl_q(2)$}, arXiv:hep-th/9304138v1.

\bibitem{citBI} 
E. Bannai and T. Ito, Algebraic Combinatorics I. Association Schemes, Benjamin/Cummings Publishing Company, 1984.

\bibitem{citTVZ}
  S. Tsujimoto, L. Vinet and A. Zhedanov,
\emph{Dunkl shift operators and Bannai-Ito polynomials}, Adv. Math. {\bf 229} (2012), 2133--2158.

\bibitem{citGVZ4} 
V.X. Genest, L. Vinet and A. Zhedanov, \emph{The Bannai-Ito polynomials as Racah coefficients of the $\sl_{-1}(2)$ algebra}, Proc. Amer. Math. Soc. {\bf 142} (2014), 1545--1560. 

\bibitem{citGVZ5} 
V.X. Genest, L. Vinet and A. Zhedanov, \emph{The Bannai-Ito algebra and a superintegrable system with reflections on the 2-sphere}, J. Phys. A: Math. Theor.  {\bf 47} (2014), 205202.. 

\bibitem{citDGV} 
H. De Bie, V.X. Genest and L. Vinet, \emph{A Dirac-Dunkl equation on $S^2$ and the Bannai-Ito algebra}, Commun. Math. Phys. {\bf 344} (2016), 447--464.
          
\bibitem{citD}
C. F. Dunkl, \emph{Differential-difference operators associated to reflection groups}, Trans. Amer. Math. Soc. {\bf311} (1989), 167-183.
 
 \bibitem{citDGTVZ}
 H. De Bie, V.X. Genest, S. Tsujimoto, L. Vinet and A. Zhedanov, \emph{The Bannai-Ito algebra and some applications}, Journal of Physics, Conference Series {\bf 597} (2015) 012001.

\end{thebibliography}
\end{document}